\documentclass{ws-ijmpd}

\usepackage{amssymb}
\usepackage{graphicx}

\begin{document}

\markboth{A.A. Andrianov, F.Cannata, A.Yu.Kamenshchik and D.Regoli}
{Phantom cosmology based on PT symmetry}

\catchline{}{}{}{}{}
%
%%%%%%%%%%%%%%%%%%%%%%%%%%%%%%%%%%%%%%%%%%%%%%%%%%%%%%%%%%%%%%%%%%%%

\title{PHANTOM COSMOLOGY BASED ON PT SYMMETRY}

% Names of the authors for the title of the paper
\author{ALEXANDER A. ANDRIANOV}

\address{
Departament d'Estructura i Constituents de la Materia and Institut de Ci\`encies del Cosmos (ICCUB)
Universitat de Barcelona, 08028, Barcelona, Spain\\
V.A. Fock Department of Theoretical Physics, Saint Petersburg State University, 198904, S.Petersburg, Russia\\andrianov@bo.infn.it}
\author{FRANCESCO CANNATA}
\address{INFN, sezione di Bologna, Via Irnerio 46, 40126 Bologna, Italy\\
cannata@bo.infn.it}
\author{ALEXANDER Yu KAMENSHCHIK}
\address{
Dipartimento di Fisica and INFN, Via Irnerio 46, 40126 Bologna,Italy\\
L.D. Landau Institute for Theoretical Physics of the Russian Academy of Sciences, Kosygin str.~2, 119334 Moscow,
Russia\\ 
kamenshchik@bo.infn.it}
\author{DANIELE REGOLI}
\address{Dipartimento di Fisica and INFN, Via Irnerio 46, 40126 Bologna, Italy\\
regoli@bo.infn.it}

\maketitle

\begin{abstract}We consider the PT~symmetric flat Friedmann  model of two scalar fields with positive kinetic terms.
While the potential of one (``normal'')  field is taken real, that  of the other field is complex. We study a
complex classical solution of the system of the two Klein-Gordon equations together with the Friedmann equation.
The solution for the normal field is real while the solution for the second field  is purely imaginary, realizing
classically the ``phantom'' behavior. The energy density and pressure are real and the corresponding geometry is
well-defined. The Lagrangian for the linear perturbations has the correct potential signs for both the fields, so
that the problem of stability does not arise. The background  dynamics is determined by an effective action
including two real fields one normal and one ``phantom''. Remarkably, the phantom phase in the cosmological
evolution is  transient and the Big Rip never occurs. Our model is contrasted to well-known quintom models, which
also include one normal and one phantom fields.
\end{abstract}


\section{Introduction}
Complex (non-Hermitian) Hamiltonians with PT~symmetry have been vigorously investigated in quantum mechanics and
quantum field theory\cite{PT}. A possibility of applications to quantum cosmology has been pointed out in
Ref.~\refcite{PT-cosm}.  In the present contribution we mainly focus attention on complex  field theory. We explore the
use of a particular  complex scalar field Lagrangian, whose  solutions of the classical equations of motion
provide us with  real physical observables and well-defined geometric characteristics.

The interest of our approach is related to its focusing on the intersection between two important fields of
research: PT~symmetric quantum theory and the cosmology of dark energy models.

As is well known the discovery of cosmic acceleration\cite{cosmic} has stimulated an intensive study of  models
of dark energy\cite{dark} responsible for the origin of this phenomenon. Dark energy is characterized by a
negative pressure whose relation to the energy density $w=p/\varepsilon$ is less than $-1/3$. Moreover, if this
relation happens to be less than $-1$ such kind of dark energy is called ``phantom'' dark energy\cite{phantom}.
The condition $w > -1$ is called the dominant energy condition. The breakdown of this condition implies the so
called super-acceleration of cosmological evolution, which in some models culminates approaching a new type of
cosmological singularity called Big Rip\cite{Rip}. While the cosmological constant ($w = -1$) is still a
possible candidate for the role of dark energy, there are observations giving  indications in favor of the models
where the equation of state parameter $w$ not only changes with time, 
but would be less than $-1$ no\cite{observe}. A standard way to introduce the phantom energy is to consider a scalar  field with the negative
sign of the kinetic energy term, usually called ``phantom'' scalar field. However such a model has been
reasonably criticized insofar as it is unstable with respect to linear perturbations\cite{crit}. Constraints on
phantom particles have been discussed in Ref.~\refcite{crit1}.

In the present paper we propose a cosmological model inspired by PT~symmetric theory\cite{PT}, choosing
potentials so that the equations of motion have classical phantom solutions for homogeneous and isotropic
universe. Meanwhile quantum fluctuations have positive energy density and this ensures the stability around a
classical background configuration. Thus our study is inspired by samples of potentials providing the real energy
spectrum bounded from below\cite{PT,andr,real,exponential,Bender-review}.

The PT~symmetric approach to the extension of quantum physics consists in the weakening of the requirement of
Hermiticity, while keeping all the physical observables real. It was shown that the  axioms of  quantum theory
are maintained if the complex extension preserves (C)PT~symmetry\cite{Bender-review}. There have also been some
attempts to apply the PT~symmetric formalism to cosmology\cite{PT-cosm,ack}. We do not enter into the discussion
about the equivalence between the PT symmetric theories and Hermitian theories with non-local Hamiltonians. The
reader is refereed to review papers in Refs.~\refcite{Bender-review,mostaf}.

We consider the complex extension of matter Lagrangians requiring the reality of all the physically measurable
quantities and the well-definiteness of geometrical characteristics. We would like to underline here, that we
work with real space-time manifolds. The attempts to use complex manifolds for studying the problem of dark
energy in cosmology were undertaken in Refs.~\refcite{complex-man}.

We start with the flat Friedmann  model of two scalar fields with positive kinetic terms. The potential of the
model is additive. One term of the potential is real, while the other is complex and PT~symmetric. We find a
classical complex solution of the system of the two Klein-Gordon equations together with the Friedmann equation.
The solution for one field  is real (the ``normal'' field) while the solution for the other field is purely
imaginary, realizing classically the ``phantom'' behavior. Moreover, the effective Lagrangian for the linear
perturbations has the correct potential signs for both the fields, so that the problem of stability does not
arise. However, the background (homogeneous Friedmann) dynamics is determined by an effective action including
two real fields one normal and one phantom. As a byproduct, we notice that the phantom phase in the cosmological
evolution is inevitably transient. The number of phantom divide line (PDL) crossings, (i.e. events such that the
ratio $w$ between pressure and energy density passes through the value $-1$)  can be only even and the Big Rip
never occurs. The avoidance of Big Rip singularity  constitutes an essential difference between our model and
well-known quintom models, including one normal and one phantom fields\cite{two-exp}. The other differences will
be discussed in more detail later.

The paper is organized as follows: in Sec.~2 we describe the cosmological model we want to analyze, together with
a brief explanation of PT~symmetric quantum mechanics; in Sec.~3 we present the results of the qualitative
analysis and of the numerical simulations for the dynamical system under consideration; conclusion and
perspectives are presented in the last section.

\section{Phantom and stability}
We shall study the flat Friedmann cosmological model described by the metric
\[
ds^2 = dt^2 - a^2(t)dl^2,
\]

where $a(t)$ is the cosmological radius of the universe. The dynamics of the cosmological evolution is
characterized by the Hubble variable
\[
h \equiv \frac{\dot{a}}{a},
\]
which satisfies the Friedmann equation
\begin{equation}
h^2 = \varepsilon, \label{Fried1}
\end{equation}
 where $\varepsilon$ is the energy density of the matter populating the universe.

We consider the matter represented by scalar fields with complex potentials. Namely, we shall try to find a
complex potential possessing the solutions of classical equations of motion which guarantee the reality of all
observables. Such an approach is inspired by the quantum theory of the PT~symmetric non-Hermitian Hamiltonians,
whose spectrum is real and bounded from below. Thus, it is natural for us to look for Lagrangians which have
consistent counterparts in the quantum theory.

Let's elucidate how the phantom-like classical dynamics arises in such Lagrangians and for this purpose choose
the one-dimensional PT symmetric potential of anharmonic oscillator $V^{(2+\epsilon)}(q) =  \lambda q^2
(iq)^\epsilon,\ 0 <\epsilon \leq 2 $ which has been proven\cite{PT,andr,real,Bender-review} to possess a real
energy spectrum bounded from below. For illustration, we restrict ourselves with $\epsilon = 2, V^{(4)}(q) =
-\lambda q^4$. The classical dynamics for {\it real} coordinates $q(t)$ offers the infinite motion with
increasing speed and energy or, in the quantum mechanical language, indicates the absence of bound states and
unboundness of energy from below. However, just there is a more consistent solution which, at the quantum level,
provides the real discrete energy spectrum, certainly, bounded from below. It has been proven, first, by means of
path integral\cite{andr} and, further on, by means of the theory of ordinary differential equations\cite{buslaev}. 
In fact, this classically ``crazy'' potential on a curve in the {\it complex} coordinate plane
generates the same energy spectrum as a two-dimensional quantum anharmonic oscillator $V^{(4)}(\vec q) =\lambda
(q_1^4 + q_2^4)/4$ with {\it real} coordinates $q_{1,2}$ in the sector of zero angular momentum\cite{andr}.
Although the superficially unstable anharmonic oscillator is well defined on the essentially complex coordinate
contour  any calculation in the style of perturbation theory (among them the semiclassical expansion) proceeds
along the contour with a fixed complex part (corresponding to a "classical" solution) and varying unboundly in
real direction. In particular, the classical trajectory for $V^{(4)}(q)$ with keeping real the kinetic, $(\dot
q)^2$ and potential, $- \lambda q^4(t)$ energies (as required by its incorporation into a cosmological scenario)
can be chosen imaginary,
$$q = i\xi,\quad \ddot \xi = -2\lambda \xi^3,\quad \dot\xi^2 = C - \lambda \xi^4,\quad C> 0 $$
which   obviously represents a bounded, finite motion with $|\xi| \leq (C/\lambda)^{1/4}$. Such a motion supports
the quasi-classical treatment of bound states with the help of Bohr's quantization. Evidently, the leading,
second variation of the Lagrangian around this solution, $q(t) = i \xi(t) + \delta q (t)$ gives a positive
definite energy,
$${\cal L}^{(2)} = p(t)\delta\dot q(t)  - H,\ H = \frac14 p^2(t) + 12 \xi^2(t) (\delta q (t))^2$$ realizing the perturbative
stability of this anharmonic oscillator in the vicinity of imaginary classical trajectory. It again reflects the
existence of positive discrete spectrum for this type of anharmonicity. However the {\it classical} kinetic
energy $-\dot\xi^2$ is negative,  i.e. it is {\it phantom-like}.

In a more general, Quantum Field Theory setting let us consider a non-Hermitian (complex) Lagrangian of a scalar
field
\begin{equation}
L = \frac12\partial_{\mu}\phi\partial^{\mu}\phi^* - V(\phi,\phi^*), \label{Lagrange}
\end{equation}
with the corresponding action,
\begin{equation}
S(\phi,\phi^*, g) =  \int \, d^4x \sqrt{-|\!|g|\!|} (L + \frac16 R(g)), \label{action}
\end{equation}
where $|\!|g|\!|$ stands for the determinant of a metric $g^{\mu\nu}$ and $R(g)$ is the scalar curvature term and
the Newton gravitational constant is normalized to $3/8\pi$ to simplify the Friedmann equations further on.

We employ potentials $V(\Phi,\Phi^*)$ satisfying the invariance condition
\begin{equation}
(V(\Phi,\Phi^*))^* = V(\Phi^*,\Phi), \label{condition}
\end{equation}
while the condition
\begin{equation}
(V(\Phi,\Phi^*))^* = V(\Phi,\Phi^*), \label{condition1}
\end{equation}
is not satisfied. This condition represents a generalized requirement of $(C)PT$ symmetry.

Let's  define two real fields,
\begin{equation}
\phi \equiv  \frac12(\Phi + \Phi^*),\quad \chi \equiv  \frac{1}{2i}(\Phi - \Phi^*) . \label{split2}
\end{equation}
Then, for example, such a potential can have a form
\begin{equation}
V(\Phi,\Phi^*) = \tilde V(\Phi+\Phi^*, \Phi-\Phi^*) = \tilde V (\phi, i\chi), \label{factor}
\end{equation}
where $\tilde V (x,y)$ is a real function of its arguments. In the last equation one can recognize the link to
the so called $(C)PT$ symmetric potentials if to supply the field $\chi$ with a discrete charge or negative
parity. When keeping in mind the perturbative stability we impose also the requirement for the second variation
of the potential to be a positive definite matrix which, in general, leads to its PT-symmetry\cite{ack}.

Here, the functions $\phi$ and $\chi$ appear as the real and the imaginary parts of the complex scalar field
$\Phi$, however, in what follows, we shall treat them as independent spatially homogeneous variables depending
only on the time parameter $t$ and, when necessary, admitting the continuation to {\it complex} values .

\section{Cosmological solution with classical phantom field}
It appears that among known PT symmetric Hamiltonians (Lagrangians) possessing the real spectrum one which is
most suitable for our purposes is that with the exponential potential. It is connected with the fact that the
properties of scalar field based cosmological models with exponential potentials are well studied\cite{power-law}. 
In particular, the corresponding models have some exact solutions providing a universe
expanding according to some power law $a(t) = a_0 t^{q}$. We shall study the model with two scalar fields and the
additive potential. Usually, in cosmology the consideration of models with two scalar fields (one normal and one
phantom, i.e. with the negative kinetic term) is motivated by the desire of describing the phenomenon of the so
called phantom divide line (PDL) crossing. At the moment of the phantom divide line crossing the equation of
state parameter $w = p/\varepsilon$ crosses the value $w = -1$ and (equivalently) the Hubble variable $h$ has an
extremum. Usually the models using two fields are called ``quintom models''\cite{two-exp}. As a matter of fact
the PDL crossing phenomenon can be described in the models with one scalar field, provided some particular
potentials are chosen\cite{weACK,weCK} or in the models with non-minimal coupling between the  scalar field and
gravity\cite{non-minimal} However, the use of two fields make all the considerations more simple and natural. In
this paper, the necessity of using two scalar fields follows from other requirements. We would like to implement
a scalar field with a complex potential to provide the effective phantom behavior of this field on some classical
solutions of equations of motion. Simultaneously, we would like to have the standard form of the effective
Hamiltonian for linear perturbations of this field. The combination of these two conditions results in the fact
the background contribution of both the kinetic and potential term in the energy density, coming from this field
are negative. To provide the positivity of the total energy density which is required by the Friedmann equation
(\ref{Fried1}) we need the other normal scalar field. Thus, we shall consider the two-field scalar Lagrangian
with the complex potential
\begin{equation}
L = \frac{\dot{\phi}^2}{2} + \frac{\dot{\chi}^2}{2} - A e^{\alpha \phi} + B e^{i\beta \chi}, \label{Lagrangian}
\end{equation}
where $A$ and $B$ are real, positive constants. This Lagrangian is the sum of two terms. The term representing
the scalar field $\phi$ is a standard one, and it can generate a power-law cosmological expansion\cite{power-law}. 
The kinetic term of the scalar field $\chi$ is also standard, but its potential is complex. In
quantum mechanics dynamical systems like the latter have been studied in the framework of PT~symmetric quantum
mechanics\cite{PT}  and, in particular, the exponential potential has been analyzed in great detail in
Ref.~\refcite{exponential}. The most important feature of this potential is that the spectrum of the corresponding
Hamiltonian is real and bounded from below, provided correct boundary conditions are assigned.

Inspired by this fact we are looking for a classical complex solution of the system, including two Klein-Gordon
equations for the fields $\phi$ and $\chi$:
\begin{equation}
\ddot{\phi} + 3 h \dot{\phi} + A\alpha e^{\alpha\phi} = 0, \label{KG}
\end{equation}
\begin{equation}
\ddot{\chi} + 3 h \dot{\chi} - iB\beta e^{i\beta\chi} = 0, \label{KG1}
\end{equation}
and the Friedmann equation
\begin{equation}
h^2 = \frac{\dot{\phi}^2}{2} + \frac{\dot{\chi}^2}{2} + A e^{\alpha \phi} - B e^{i\beta \chi}. \label{Friedmann}
\end{equation}
The classical solution which we are looking for should provide the reality and positivity of the right-hand side
of the Friedmann equation (\ref{Friedmann}). The solution where the scalar field $\phi$ is real, while the scalar
field $\chi$ is purely imaginary
\begin{equation}
\chi = i\xi,\quad\xi\ {\rm real}, \label{xi}
\end{equation}
uniquely satisfies this condition. Moreover, the Lagrangian (\ref{Lagrangian}) evaluated on this solution is real
as well. This is remarkable because on homogeneous solutions the Lagrangian coincides with the pressure, which
indeed should be real.

Substituting the equation(\ref{xi})  into the Friedmann equation (\ref{Friedmann}) we shall have
\begin{equation}
h^2 = \frac{\dot{\phi}^2}{2} - \frac{\dot{\xi}^2}{2} + A e^{\alpha \phi} - B e^{-\beta \xi}. \label{Friedmann1}
\end{equation}
Hence, effectively we have the Friedmann equation with two fields: one ($\phi$) is a standard scalar field, the
other ($\xi$) has the phantom behavior, as we pointed out above. In the next section we shall study the
cosmological dynamics of the (effective) system, including (\ref{Friedmann1}), (\ref{KG}) and
\begin{equation}
\ddot{\xi} + 3 h \dot{\xi} - B\beta e^{-\beta\xi} = 0. \label{KG2}
\end{equation}

The distinguishing feature of such an approach rather than the direct construction of phantom Lagrangians becomes
clear when one calculates the linear perturbations  around the classical solutions. Indeed the second variation
of the action for the field $\chi$ gives the quadratic part of the effective Lagrangian of perturbations:
\begin{equation}
L_{eff} = \frac12 \dot{\delta \chi}^2 -\frac12 (\vec{\nabla}(\delta \chi))^2 - B\beta^2 e^{i\beta\chi_0} (\delta\chi)^2 , \label{perturb}
\end{equation}
where $\chi_0$ is a homogeneous purely imaginary solution of the dynamical system under consideration
and $\vec{\nabla}$ stands for spatial gradients . It is easy
to see that on this solution, the effective Lagrangian (\ref{perturb}) will be real and its potential term has a
sign providing the stability of  the background solution with respect to linear perturbations as the related
Hamiltonian is positive,
\begin{equation}
H_{eff}^{(2)} = \frac12 {\delta \pi}^2  + \frac12 (\vec{\nabla}(\delta \chi))^2 + B\beta^2 e^{-\beta\xi_0} (\delta\chi)^2 ,\quad \delta \pi
\Leftrightarrow \delta\dot\chi . \label{perturbham}
\end{equation}

The Hamiltonian of the metric perturbations should be naturally added to the above formulae.
This Hamiltonian includes two types of terms: bilinear terms in metric perturbations and mixed terms
, including both the metric and  scalar field 
perturbations. 
The part of the second-order action, including terms bilinear in the metric perturbations can be represented in the 
following form \cite{pert1,pert2}:
\begin{eqnarray}
&&S_{g,\ g}^{(2)} = \int d^4 x \Big\{ \frac{1}{12}  A^2(\tau) \Big[ \frac12 h^{\mu\nu,\rho} h_{\mu\nu,\rho}
-   h^{\mu\nu}_{,\mu} h^\rho_{\nu,\rho}
+ h^{\mu\nu}_{,\nu} h_{,\mu}
- \frac{1}{2} h_{,\mu} h^{,\mu}\Big] 
\nonumber\\
&&+ A(\tau) A'(\tau) \Big[\frac16 ( h^{\mu\nu} h_{0\nu,\mu} - h^{\mu\nu}_{,\mu} h_{0\nu}) + \frac{1}{12} (h^{\mu 0} h_{,\mu}-h^{\mu 0}_{,\mu} h)\Big]\\
&&- \frac{1}{12}(A'^2(\tau) +A(\tau)A''(\tau)) h^{00} h \Big\},
\label{pert1}
\end{eqnarray}
where the ``prime'' means the derivative with respect to conformal time $\tau$ and the background metrics has the form
\begin{equation*}
ds^2 = A^2(\tau)(d\tau^2 - dl^2),
\end{equation*}
while the full metric is 
\begin{equation*}
g_{\mu\nu}(x) = A^2(\tau) (\eta_{\mu\nu} + h_{\mu\nu}(x)).
\end{equation*}

The above expression (\ref{pert1}) was obtained by the proper integration by parts and taking into account 
the equations of motion \cite{pert1,pert2}. 
The only dangerous term here is the last ``massive'' term, which can be removed by the choice of gauge fixing condition
\begin{equation}
h^{00} = 0.
\label{gauge-fix}
\end{equation}
The second-order action, including mixed terms (the metric perturbations and the perturbations of the field $\chi$) 
has the following form:
\begin{equation}
S^{(2)}_{g,\,\chi}=\int d ^{4}x A^2(\tau)\left\{\chi'\left(\frac12 h_{,0}-h^{0\mu}_{\;\; ,\mu}\right)-\left(\chi''+2\frac{A'(\tau)}{A(\tau)}\chi'\right)h^{00}\right\}\delta\chi,
\label{pert2}
\end{equation}
where again we have used the equation of motion (the Klein-Gordon equation for the field $\chi$.
This term (\ref{pert2})
appears  annoying because it contains the first derivative of 
the term $B e^{i\beta \chi}$, which is imaginary.   
However, one can show that by a proper choice of the gauge condition this imaginary term can be eliminated. 
Indeed, the gauge condition (\ref{gauge-fix}) eliminate from (\ref{pert2}) the term, proportional to $h^{00}$. 
Then remains the term proportional to $\chi'$ multiplied by the linear combination of the metric perturbations.
This combination can be annihilated by the following choice of the second gauge condition:
\begin{equation}
\frac12 h_{,0}-h^{0\mu}_{\;\; ,\mu} = 0.
\label{gauge-fix1}
\end{equation}
This condition is compatible with the first gauge condition (\ref{gauge-fix}).  
Thus, we have shown that the stability with respect to perturbations in our model  can be proved by fixing a proper 
gauge condition. 
%Let us explain it in more detail. The mixed variation of the action with respect to $\delta\chi$ %and 
%$h_{\mu\nu} \equiv \delta g_{\mu\nu}$, where Greek indices stand for spacetime coordinate is 
%\begin{eqnarray}
%&&\delta^2 S = \int d^4x \sqrt{-g} %h_{\mu\nu}\left(\frac12g^{\mu\nu}\left(\partial_{\alpha}\delta\chi %\partial_{\beta}\chi^{(0)}g^{\alpha\beta} -
%V'(\chi^{(0)})\delta\chi\right)\right.\nonumber \\
%&&\left.-\partial^{\mu}\delta\chi\partial^{\nu}\chi^{(0)}\right),
%\label{variation}
%\end{eqnarray}
%where $\chi^{(0)}$ is the purely imaginary solution of the classical equation of motion (\ref{xi}).
%Integrating by parts and using the classical equation of motion (\ref{KG2}) we transfrom the second variation (\ref{variation})
%to 
%\begin{eqnarray}
%&&\delta^2 S = \int d^4x \sqrt{-g}\delta\chi \left( h_{00} \chi^{(0)}_{,00} \right.\nonumber \\
%&&\left. + \chi^{(0)}_{,0}\left(\nabla_{\mu}h^{\mu 0} -\frac12h_{,0} + %\frac{\dot{a}}{a}(h-h_{00})\right)\right),
%\label{variation1}
%\end{eqnarray}
%where $h = h^\mu_\mu$, commas stand for partial derivative and $\nabla_\mu$ is the covariant %derivative operator.
%The expression (\ref{variation1}) is purely imaginary because $\chi^{(0)}$ is purely imaginary (see (\ref{xi})). 
%Let us show that this structure can be eliminated by the gauge transformation
%\begin{equation}
%h_{\mu\nu} \rightarrow h_{\mu\nu} + \zeta_{\mu}
%\end{equation}
%\begin{eqnarray}
%&&\sqrt{-g}\delta\chi\left(\chi^{(0)}_{,0}(\nabla_\mu h^{\mu 0} - frac12 h^{,0}+\box\xi^0\right) + %h_{00}\chi^{(0)}_{,00} \nonumber\\
%&&
%\end{eqnarray}

Let us list the main differences between our model and quintom models, using two fields (normal scalar and
phantom) and exponential potentials\cite{two-exp}. First, we begin with two normal (non-phantom) scalar fields,
with normal kinetic terms, but one of these fields is associated to a complex (PT~symmetric) exponential
potential. Second, the (real) coefficient multiplying this exponential potential is negative. Third, the
background classical solution of the dynamical system, including two Klein-Gordon equations and the Friedmann
equation,  is such that the second field is purely imaginary, while all the geometric characteristics are
well-defined. Fourth, the interplay between transition to the purely imaginary solution of the equation for the
field $\chi$ and the negative sign of the corresponding potential provides us with the effective Lagrangian for
the linear perturbations of this field which have correct sign for both the kinetic and potential terms: in such
a way the problem of stability of the our effective phantom field is resolved. Fifth, the qualitative analysis of
the corresponding differential equations, shows that in contrast to the quintom models in our model the Big Rip
never occurs. The numerical calculations confirm this statement.

In the next section we shall describe the cosmological solutions for our system of equations.

\section{Cosmological evolution}
First of all notice that our dynamical system  permits the existence of cosmological trajectories which cross
PDL. Indeed, the crossing point is such that the time derivative of the Hubble parameter
\begin{equation}
\dot{h} = -\frac32(\dot{\phi}^2 - \dot{\xi}^2) \label{PDL}
\end{equation}
is equal to zero. We always can choose $\dot{\phi} = \pm \dot{\xi}$, at  $t = t_{PDL}$ provided the values of the
fields $\phi(t_{PDL})$ and $\xi(t_{PDL})$ are chosen in such a way, that the general potential energy $A
e^{\alpha \phi} - B e^{-\beta \xi}$ is non-negative. Obviously, $t_{PDL}$ is the moment of  PDL crossing.
However, the event of the PDL crossing cannot happen only once. Indeed, the fact that the universe has crossed
phantom divide line means that it was in effectively phantom state before or after such an event, i.e. the
effective phantom field $\xi$ dominated over the normal field $\phi$. However, if this dominance lasts for a long
time it implies that non only the kinetic term $-\dot{\xi}^2/2$ dominates over the kinetic term $\dot{\phi}^2/2$
but also the potential term $-B\exp(-\beta\xi)$ should dominate over $A\exp(\alpha\phi)$; but it is impossible,
because contradicts to the Friedmann equation (\ref{Friedmann1}). Hence, the period of the phantom dominance
should finish and one shall have another point of PDL crossing. Generally speaking, only the regimes with even
number of PDL crossing events are possible. Numerically, we have found only the cosmological trajectories with
the double PDL crossing. Naturally, the trajectories which do not experience PDL crossing at all also exist and
correspond to the permanent domination of the normal scalar field. Thus, in this picture,  there is no place for
the Big Rip singularity as well, because such a singularity is connected with the drastically dominant behavior
of the effective phantom field, which is impossible as was explained above. The impossibility of approaching the
Big Rip singularity can be argued in a more rigorous way as follows. Approaching the Big Rip, one has a growing
behavior of the scale factor $a(t)$ of the type $a(t) \sim (t_{BR}-t)^{-q}$, where $q >0$. Then the Hubble
parameter is
\begin{equation}
h(t) = \frac{q}{t_{BR} -t} \label{Hubble3}
\end{equation}
and its time derivative
\[
\dot{h}(t) = \frac{q}{(t_{BR}-t)^2}.
\]
Then, according to Eq.~(\ref{PDL}),
\begin{equation}
\frac12\dot{\xi}^2 - \frac12\dot{\phi}^2 = \frac{q}{3(t_{BR}-t)^2}. \label{BR}
\end{equation}
Substituting Eqs.~(\ref{BR}), (\ref{Hubble3}) into the Friedmann equation (\ref{Friedmann1}), we come to

\[
\frac{q}{(t_{BR}-t)^2}=\frac{q}{3(t_{BR}-t)^2}+Ae^{\alpha\phi}-Be^{-\beta\xi}.
\]

In order for this to be satisfied and consistent, the potential of the scalar field $\phi$ should behave as
$1/(t_{BR}-t)^2$. Hence the field $\phi$ should be
\begin{equation}
\phi = \phi_0 -\frac{2}{\alpha} \ln (t_{BR} - t), \label{log}
\end{equation}
where $\phi_0$ is an arbitrary constant. Now substituting Eqs.~(\ref{Hubble3}) and (\ref{log}) into the
Klein-Gordon equation for the scalar field $\phi$ (\ref{KG}), the condition of the cancelation of the most
singular terms in this equation which are proportional to $1/(t_{BR}-t)^2$ reads
\begin{equation}
2 + 6q + A\alpha^2 \exp(\alpha{\phi_0}) = 0. \label{cond}
\end{equation}
This condition cannot be satisfied because all the terms in the left-hand side of Eq.~(\ref{cond}) are positive.
This contradiction demonstrates that it is impossible to reach the Big Rip.

\begin{figure}[htp]\centering
\begin{minipage}{.5\textwidth}\centering
\includegraphics[scale=.8]{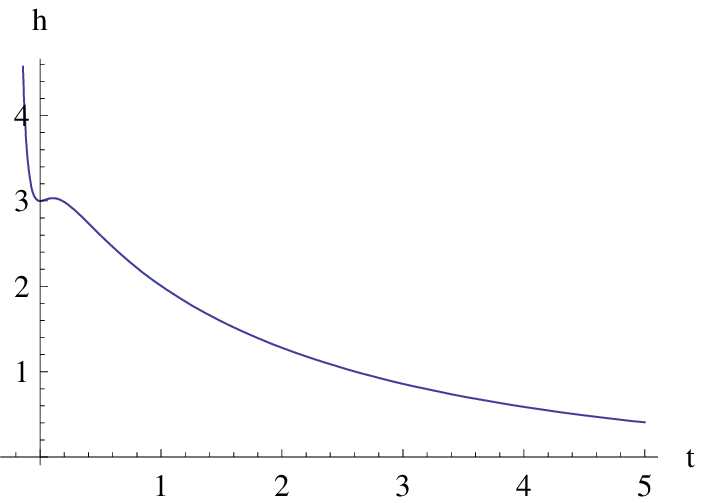}
\end{minipage}%
\begin{minipage}{.5\textwidth}\centering
\includegraphics[scale=0.8]{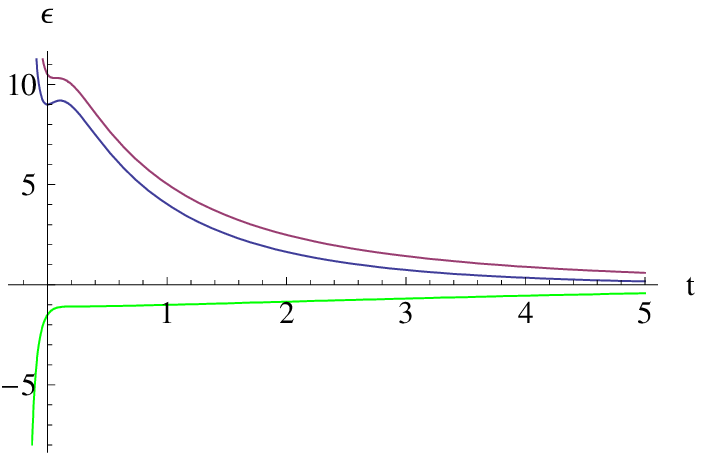}
\end{minipage}
\caption{(left) Plot of the Hubble parameter representing the cosmological evolution. The evolution starts from a
Big Bang-type singularity and goes through a transient phase of superaccelerated expansion (``phantom era''),
which lies between two crossings of PDL (when the derivative of $h$ crosses zero). Then the universe expands
infinitely. (right) Plots of the total energy density (blue), and of the energy density of the normal field
(purple) and of the phantom one (green).}\label{fig1} 
\end{figure}

\begin{figure}[htp]\centering
\begin{minipage}{.5\textwidth}\centering
\includegraphics[scale=.8]{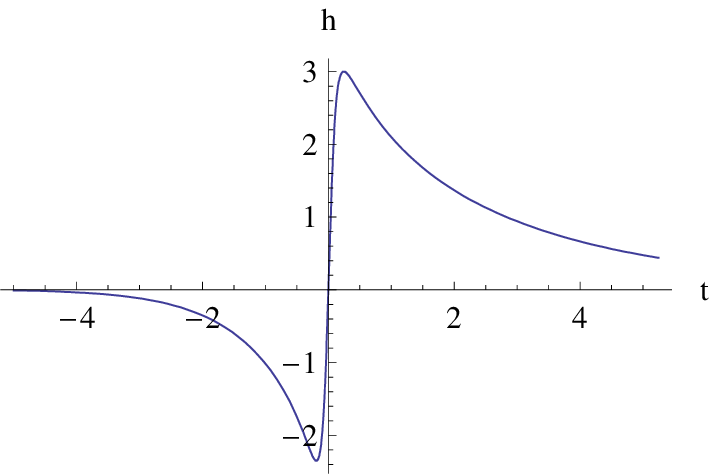}
\end{minipage}%
\begin{minipage}{.5\textwidth}\centering
\includegraphics[scale=.8]{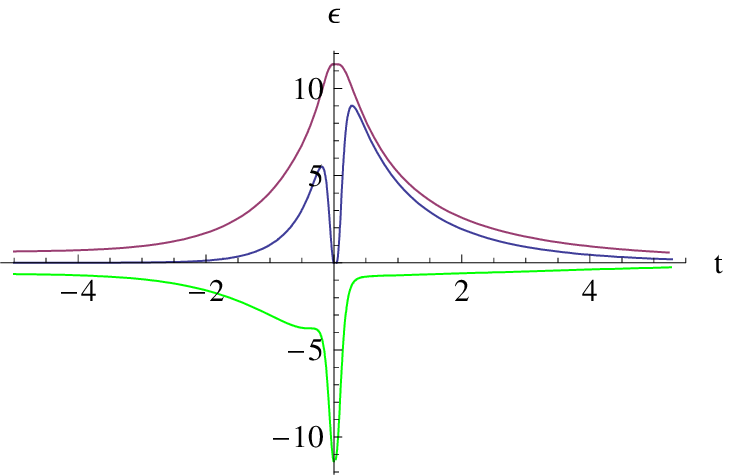}
\end{minipage}
\caption{(left) The evolution starts with a contraction in the infinitely remote past. At the point first PDL
crossing the contraction becomes superdecelerated and  turns in a superaccelerated expansion when $h$ crosses
zero. The second PDL crossing ends the ``phantom era''; the decelerated expansion continues till the universe
begins contracting. In a finite time a Big Crunch-type singularity is reached. (right) Plots of the total energy
density (blue), and of the energy density of the normal field (purple) and of the phantom one
(green).}\label{fig2}
\end{figure}

\begin{figure}[htp]\centering
\begin{minipage}{.5\textwidth}\centering
\includegraphics[scale=.8]{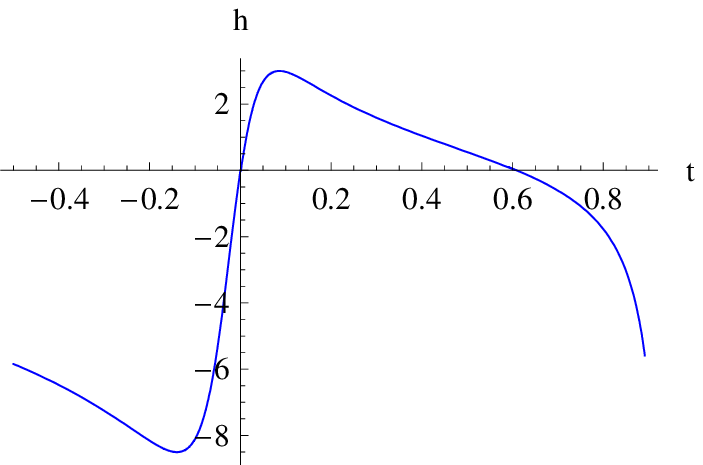}
\end{minipage}%
\begin{minipage}{.5\textwidth}\centering
\includegraphics[scale=.8]{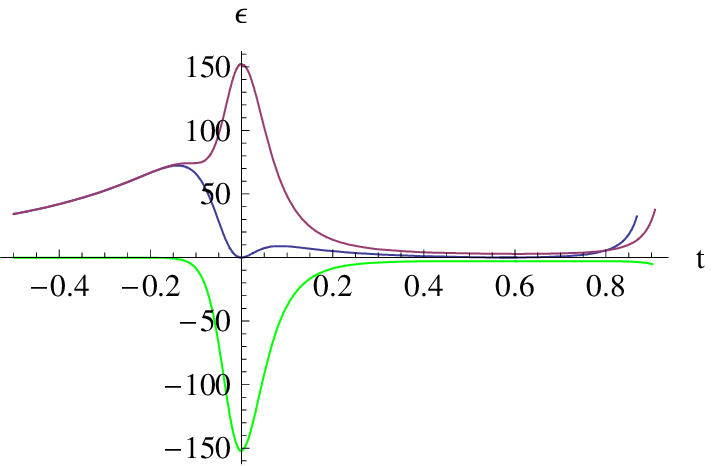}
\end{minipage}
\caption{(left) The cosmological evolution begins with a contraction in the infinitely remote past. with the
first PDL crossing the contraction becomes superdecelerated until the universe stops ($h=0$) and starts
expanding. With the second crossing the ``phantom era'' ends and the expansion continues infinitely. (right)
Plots of the total energy density (blue), and of the energy density of the normal field (purple) and of the
phantom one (green).}\label{fig3}
\end{figure}

\begin{figure}[htp]\centering
\begin{minipage}{.5\textwidth}\centering
\includegraphics[scale=.8]{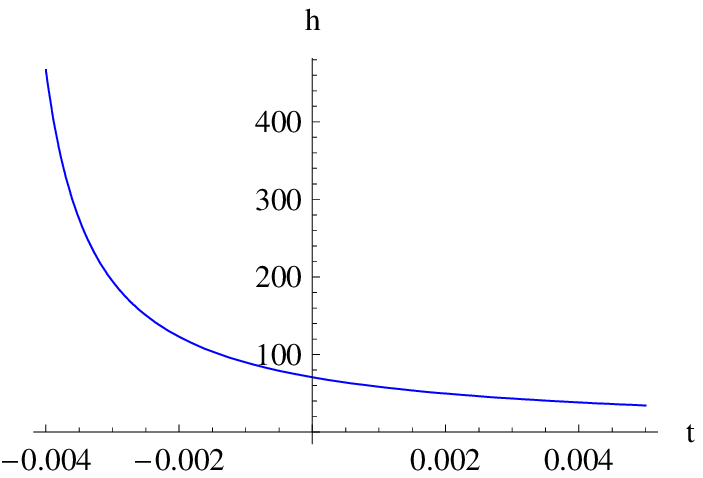}
\end{minipage}%
\begin{minipage}{.5\textwidth}\centering
\includegraphics[scale=.8]{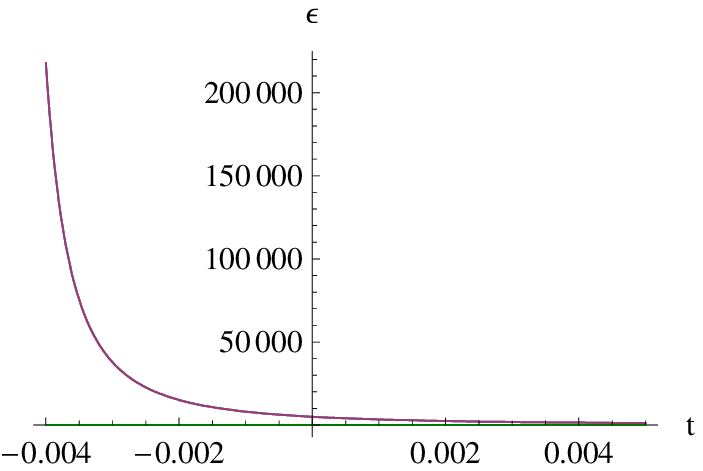}
\end{minipage}
\caption{(left) Evolution from a Big Bang-type singularity to an infinite expansion, without any crossing of PDL.
This evolution is thus guided by the ``normal'' field $\phi$. (right) Plots of the total energy density (blue),
and of the energy density of the normal field (purple) and of the phantom one (green). Notice that the energy
density of the phantom field (green) is very close to zero, thus the total energy density is mainly due to the
standard field.}\label{fig4}
\end{figure}

Now we would like to describe briefly some examples of cosmologies contained in our model, deduced by numerical analysis of the system of equations of motion.\\
In Fig.~\ref{fig1} a double crossing of PDL is present. The evolution starts from a Big Bang-type singularity and
goes through a transient phase of super-accelerated expansion (``phantom era''), which lies between two crossings
of PDL. Then the universe undergoes an endless expansion. In the right plot we present the time evolution of the
total energy density and of its partial contributions due to the two fields, given by the equations
\begin{eqnarray*}
&&\varepsilon_\phi = \frac{1}{2}\dot\phi^2+Ae^{\alpha\phi},\\
&&\varepsilon_\xi = -\frac{1}{2}\dot\xi^2-Be^{\beta\xi},\\
&&\varepsilon =\varepsilon_\phi+\varepsilon_\xi,
\end{eqnarray*}
which clarify the roles of the two fields in driving the cosmological evolution.\\
The evolution presented in Fig.~\ref{fig2} starts with a contraction in the infinitely remote past. Then the contraction becomes superdecelerated and  turns later in a superaccelerated expansion . With the second PDL crossing the ``phantom era'' ends; the decelerated expansion continues till the universe begins contracting. After a finite time a Big Crunch-type singularity is encountered. From the right plot we can clearly see that the ``phantom era'' is indeed characterized by a bump in the (negative) energy density of the phantom field.\\
In Fig.~\ref{fig3} the cosmological evolution again begins with a contraction in the infinitely remote past. Then the universe crosses PDL: the contraction becomes superdecelerated until the universe stops  and starts expanding. Then the "phantom era" ends and the expansion is endless.\\
In Fig.~\ref{fig4} the evolution from a Big Bang-type singularity to an eternal expansion is shown. The phantom
phase is absent. Indeed the phantom energy density is almost zero everywhere.

\subsection{The cosmological dynamics in the presence of dust like matter}
Trying to make our model more realistic we can try to describe which changes does it undergo in the presence of 
some matter. The natural candidate for the role of matter is a dust-like perfect fluid with the vanishing pressure.
In this case the Friedmann equation (\ref{Friedmann1}) acquires the form
\begin{equation}
h^2 = \frac{\dot{\phi}^2}{2} - \frac{\dot{\xi}^2}{2} + A e^{\alpha \phi} - B e^{-\beta \xi} + \frac{C_0}{a^3},
\label{Friedmann10}
\end{equation}
while the second Friedmann equation (\ref{PDL}) becomes
\begin{equation}
\dot{h} = -\frac32\left(\dot{\phi}^2 - \dot{\xi}^2+\frac{C_0}{a^3}\right), \label{PDL10}
\end{equation}
where the constant $C_0$ characterizes the quantity of the dust-like matter in a universe under consideration.

Obviously at the early stage of the cosmological evolution, this dust contribution into the Friedmann equations dominates 
over dark energy contribution coming from our scalar fields. So, at early stage of the evolution (close to the Big Bang epoch)
the effects like phantomization of the matter content of the universe and phantom divide line crossing are impossible.
They are postponed until the time when the expansion of the universe is so large that this additional term becomes insignificant. However, one can say that its presence even underlines some peculiar effects of the considered model.
Indeed, it makes stronger the non-phantom part of the matter content of the universe and hence return to the normal non-phantom 
case occurs more quickly. Naturally, the Big Rip singularity is avoided in this model.

\section{Conclusion}
As was already said above  the data are compatible with the presence of the phantom energy, which, in turn, can
be in a most natural way realized by the phantom scalar field with a negative kinetic term. However, such a field
suffers from the instability problem, which makes it vulnerable. Inspired by the development of PT~symmetric
quantum theory\cite{Bender-review} we introduced the PT~symmetric two-field cosmological model where both the
kinetic terms are positive, but the potential of one of the fields is complex. We studied a classical background
solution of two Klein-Gordon equations together with the Friedmann equation, when one of this fields (normal) is
real while the other is purely imaginary. The scale factor in this case is real and positive just like the energy
density and the pressure. The background dynamics of the universe is determined by two effective fields - one
normal and one phantom, while the Lagrangian of the linear perturbations has the correct sign of the mass term.
Thus, so to speak, the quantum normal theory is compatible with the classical phantom dynamics and the problem of
instability is absent.
%Quantization will involve the effective Lagrangian (\ref{perturb}), thus particles associated to the field $\chi$ will %have positive energy. Therefore the constraints arising from the analysis \cite{crit1} of negative energy particles do %not apply to our scheme. Indeed we have obtained a realization of phantom classical (background) fields without phantom %particles.

As a byproduct of the structure of the model, at variance with quintom models, the phantom dominance era is transient, the number of the phantom divide line crossings is even and the Big Rip singularity is excluded.
In all our numerical simulations we have seen only two phantom divide line crossings. However, one cannot exclude that 
at some special initial conditions, or, in the models of the kind  considered in the present paper, but 
with more complicated potentials, it is possible to see more than two phantom divide line crossings. It would be especially interesting to find the potentials or/and initial conditions providing an oscillatorial behaviour with infinite phantom divide line crossings.

\section*{Acknowledgments} This work was partially supported by Grants RFBR  08-02-00923  and  LSS-4899.2008.2. 
The work of A.A. was supported by grants FPA2007-66665-C02-01, 2009SGR502,
Consolider-Ingenio 2010 Program CPAN (CSD2007-00042), by Grants RFBR 09-02-00073-a,
09-01-12179-ofi-m  and Program RNP2009-2.1.1/1575.

\end{document}